\documentclass[aps,pra,amssymb,superscriptaddress,10pt,tightenlines,twocolumn]{revtex4-1}

\usepackage[english]{babel}
\usepackage{graphicx}
\usepackage{dcolumn}
\usepackage{bm}
\usepackage{color}
\usepackage{subfigure}
\usepackage{amsfonts}
\usepackage{amsmath}
\usepackage{amssymb}
\usepackage{hyperref}
\usepackage{esint}

\def\bra#1{\langle {#1} |}
\def\ket#1{| {#1} \rangle}

\def\ee{\mathrm{e}}
\def\dd{\mathrm{d}}
\def\ii{\mathrm{i}}

\def\H{\hat{H}}
\def\b{\hat{b}}
\def\c{\hat{c}}

\newcommand{\unibari}{Dipartimento di Fisica and MECENAS - Universit\`{a} di Bari, I-70126 Bari, Italy}
\newcommand{\infn}{INFN, Sezione di Bari, I-70126 Bari, Italy}
\newcommand{\ino}{{Istituto Nazionale di Ottica (INO-CNR), I-50125 Firenze, Italy}}
\newcommand{\ictp}{The Abdus Salam ICTP, I-34151 Trieste, Italy}

\begin{document}

\author{Francesco V. Pepe}
\affiliation{\unibari}
\affiliation{\infn}

\author{Paolo Facchi}
\affiliation{\unibari}
\affiliation{\infn}

\author{Zeinab Kordi}
\affiliation{\ictp}
\affiliation{\infn}

\author{Saverio Pascazio}
\affiliation{\unibari}
\affiliation{\infn}
\affiliation{\ino}

\title{Nonexponential decay of Feshbach molecules}

\begin{abstract}
We analyze the temporal behavior of the survival probability of an unstable $^6$Li Feshbach molecule close to the BCS-BEC crossover. We find different instances of nonexponential decay as the magnetic field approaches the resonance value, at which the molecule becomes stable. We observe a transition from an exponential decay towards a regime dominated by a stretched-exponential law. 
\end{abstract}

\maketitle

\section{Introduction}
The decay of an unstable system is commonly associated with an exponential law, which classically derives from the assumption that the survival probability decreases at a rate proportional to its value at that time. However, the quantum evolution of an unstable state, governed by the Schr\"odinger equation, features deviations from the exponential law~\cite{gamow28,WW1,WW2,strev}. On one hand, at short times, unless the initial state has infinite energy variance, the survival probability is quadratic~\cite{zenoreview1,zenoreview2,artzeno}. On the other hand, the exponential regime cannot last indefinitely if a ground state exists~\cite{Khalfin57,Khalfin58,Exner85,Muga09}, and a slower (typically, power-law) decay takes over (see, however,~\cite{Burgarth17}).  In atomic and nuclear physics decays, the aforementioned deviations are usually so small to be practically unobservable, the exponential law having been verified with a high degree of accuracy (see e.g.~\cite{expon1,expon2}). The initial quadratic regime of decay has been however experimentally confirmed on a number of carefully controlled physical systems~\cite{Wilkinson,FGR,Ketterle,Raimond2010,Firenze2014,Signoles2014,Reichel,Wineland}, while the (more elusive) power-law decay at long times has been experimentally confirmed only very recently~\cite{ladder,mdpicrespi}.

The presence of strong coupling and/or a structured spectrum can induce peculiar deviations from the expected exponential law. In this article, we will consider the decay of a weakly bound Feshbach molecule. A Feshbach resonance in the collision of two atoms, either identical or not, in a given internal configuration (the \textit{open channel}) occurs whenever the energy of the scattering state is close to the energy of a molecular bound state in a different internal state (the \textit{closed channel})~\cite{grimm}. In this condition, the $s$-wave scattering length diverges, with the molecular state being energetically favored with respect to the unbound atomic state on the side of positive divergence. We will be interested, in particular, in magnetic resonances, whose control parameter is an external magnetic field. Feshbach resonances give rise to a variety of physical phenomena, including the crossover, at thermal equilibrium, between a BCS state of atomic (fermionic) Cooper pairs and a Bose-Einstein condensation of weakly bound bosonic molecules~\cite{crossover1,crossover2,crossover3,crossover4}, and represents a powerful instrument to investigate the atomic structure and interaction potentials~\cite{moleculeapp1,moleculeapp2}. The possibility to obtain a stable molecular state by adiabatically varying the magnetic field in presence of a resonance~\cite{conversion1,conversion2,conversion3,conversion4}, as well as the effects of atom-molecule coherence in the evolution of the system~\cite{AMcoherence1,AMcoherence2}, have been extensively investigated.

In this Article, we will consider deviations from exponential decay laws in systems close to Feshbach resonances. As explained above, this is a general and fertile arena, that enables one to study a plethora of interesting fundamental physical effects. Here, for the sake of concreteness, we will focus on the time evolution of an unstable $^6$Li molecular state~\cite{lithium1995,lithium}, associated to the resonance at $543.25\,\mathrm{G}$ magnetic field~\cite{grimm,lithium}. This choice is motivated by the fact that the considered resonance is closed-channel dominated, in the sense that, outside the small range of $0.1\,\mathrm{G}$ from the resonance, the molecular state has negligible hybridization with the atomic sector. In characterizing the molecule decay, we will unveil the emergence of a stretched-exponential regime~\cite{stretched2}, that becomes dominant as the resonance is approached. 

Dynamics yielding the appearance of stretched exponentials are typical of a number of different phenomena, both in statistical phenomena \cite{stretched2}, glassy dynamics \cite{stretched1}, and in the context of 1D Bose Gas, when the low energy physics is well described by a Tomonaga-Luttinger liquid~\cite{Citro}. Interestingly, in the latter case the single-particle correlation function also displays a power-law behavior. Although the coefficients characterizing the stretched exponential can be different, the underlying phenomena involves the presence of different, competing dynamics and lifetimes.

Our Article is organized as follows.
In Section~\ref{sec:hamiltonian} we introduce the model Hamiltonian and characterize the initial state. In Section~\ref{sec:form} we discuss the relation between the properties of the Hamiltonian and the molecular state with the form factors of atom-molecule interactions. In Section~\ref{sec:morse} the interatomic potential in the molecular state is fitted by a Morse potential, providing semi-analytic results. In Section~\ref{sec:self} we introduce the resolvent formalism to characterize the time evolution of the molecule. In Section~\ref{sec:poles} we characterize the decay rate of the molecule, in connection with the analytic structure of the propagator of the initial state. In Section~\ref{sec:time} we discuss the different relevant regimes in the time evolution of the molecule. Finally, in Section~\ref{sec:conc}, we summarize our results and comment on possible perspectives.

\section{Hamiltonian and initial state}\label{sec:hamiltonian}

We will consider the boson-fermion Hamiltonian
\cite{bosefermi1,bosefermi2,bosefermi3}
\begin{equation}\label{eq:hamilt}
\H = \H_0 + \H_{\mathrm{AM}} + \H_{\mathrm{F}},
\end{equation}
with
\begin{equation}
\H_0 = \sum_{\bm{p}} \sum_{\sigma=\uparrow,\downarrow} \frac{p^2}{2
m} \c_{\bm{p}, \sigma}^{\dagger} \c_{\bm{p}, \sigma}^{\,} +
\sum_{\bm{q}} \left( \frac{q^2}{4m} + E_B \right)
\b_{\bm{q}}^{\dagger} \b_{\bm{q}}^{\,}
\end{equation}
describing the dynamics of free fermionic atoms of mass $m$ and bosonic molecules of mass $2m$.  $E_B$ is the molecular binding energy, that is the difference between the energy of the static resonant molecular state (``closed channel'') and the continuum threshold for the free atom pair, and is an approximately linear function of the magnetic field~\cite{grimm}. The field operators $\b_{\bm{q}}$ and
$\c_{\bm{p},\sigma}$, which satisfy the canonical (anti)commutation relations
\begin{align}
[\b_{\bm{q}},\b_{\bm{q}'}]=0, \quad &
[\b_{\bm{q}}^{\,},\b_{\bm{q}'}^{\dagger}] =
\delta_{\bm{q},\bm{q}'} ,  \\ \{ \c_{\bm{p},\sigma},
\c_{\bm{p}',\sigma'} \}=0, \quad & \{ \c_{\bm{p},\sigma},
\c_{\bm{p}',\sigma'}^{\dagger} \} = \delta_{\bm{p} \bm{p}'}
\delta_{\sigma \sigma'},
\end{align}
act on the Fock space of the atom-molecule system. The atoms are characterized by two possible internal states (pseudospins), denoted by $\uparrow$ and $\downarrow$. 

The interaction Hamiltonian $\H_{\mathrm{AM}}$  describes the transitions between a pair of atoms with opposite pseudospin and a molecule, preserving the total momentum:
\begin{equation}
\H_{\mathrm{AM}} = \sum_{\bm{K}, \bm{p}} \left( G(\bm{p}) \,
\b_{\bm{K}}^{\dagger} \c_{-\bm{p} + \bm{K}/2 ,\downarrow}^{\,}
\c_{\bm{p} + \bm{K}/2 ,\uparrow}^{\,} + \mathrm{H.c.}  \right).
\label{eq:HAMdef}
\end{equation}
The coupling $G(\bm{p})$ between the molecule and the
atom pair is assumed to be independent of
the total momentum $\bm{K}$ by Galilean invariance, and reads,
according to the second-quantization prescription~\cite{fetter}
\begin{equation}\label{eq:amcoupling}
G(\bm{p}) = \bra{ \psi_{\mathrm{M},0 } } H_{\mathrm{int}}
\ket{ \bm{p} \uparrow, - \bm{p} \downarrow },
\end{equation}
where
\begin{eqnarray}
\ket{ \psi_{\mathrm{M},0 } } & = & \b_{\bm{0}}^{\dagger}
\ket{0},
\label{eq:instate}
\\ 
\ket{ \bm{p} \uparrow, - \bm{p} \downarrow
} & = & \c_{\bm{p},\uparrow}^{\dagger} \c_{-\bm{p}, \downarrow}^{\dagger} \ket{0},
\label{eq:relmotdef}
\end{eqnarray}
with $\ket{0}$ the vacuum of the Fock space and
$H_{\mathrm{int}}$ the (first-quantization) Hamiltonian that couples
the atomic and molecular sectors. We will later characterize the function $G(\bm{p})$ for our case study.

The last term in~\eqref{eq:hamilt} describes two-body
interactions between atoms in different internal states, through an interatomic potential that depends only on their relative position:
\begin{equation}
\label{eq:quartic}
\H_{\mathrm{F}} \! = \!\! \sum_{\bm{p} , \bm{p}' , \bm{q}} \! U(\bm{p}-\bm{p}')
\c_{\bm{p} + \bm{q}/2 ,\uparrow}^{\dagger} \c_{-\bm{p} + \bm{q}/2
,\downarrow}^{\dagger} \c_{-\bm{p}' + \bm{q}/2 ,\downarrow}^{\,}
\c_{\bm{p}' + \bm{q}/2 ,\uparrow}^{\,}.
\end{equation}
Since, at low-energy, the $s$-wave contribution dominates two-body scattering, we will neglect interactions between atoms in the same internal state, which appear for higher-order partial waves.

Our goal is to characterize the evolution of the initial one-molecule state (\ref{eq:instate})
through its survival probability at time $t$
\begin{equation}
P(t) = |\bra{\psi_{\mathrm{M},0}} \ee^{-\ii \frac{t}{\hbar} \hat{H}} \ket{\psi_{\mathrm{M},0}}|^2 .
\end{equation}
Notice that in experiments with ultracold gases, a molecular condensate can be prepared, represented by a many-body state $\ket{\Psi_0}$ with
\begin{equation}
\bra{\Psi_0} \b_0^{\dagger} \b_0^{\,} \ket{\Psi_0} = N ,
\end{equation} 
where the typical values of the particle number are $N\simeq 10^4$--$10^7$~\cite{bloch}. We will assume that, at least in the initial part of the evolution (when the molecular survival probability $\gtrsim 0.5$), which is of interest for this Article, the effects of intermolecular scattering, mediated by the atom pairs emitted by different molecules, are suppressed. Thus, the evolved condensate fraction
\begin{equation}
N(t)=\bra{\Psi_0} \ee^{\ii \frac{t}{\hbar}\hat{H}} \b_0^{\dagger} \b_0^{\,} \ee^{-\ii \frac{t}{\hbar}\hat{H}} \ket{\Psi_0}
\end{equation}
will be proportional to the survival probability~$P(t)$ with good approximation.

\section{Form factor of the atom-molecule interaction}\label{sec:form}

The atom-molecule form factor is determined by the first-quantization interaction Hamiltonian $\H_{\mathrm{int}}$ and the details of the molecular state~$\ket{ \psi_{\mathrm{M},0 } }$. In order to properly analyze these aspects, we will consider the $s$-wave Feshbach resonance of $^6$Li ($m\simeq 10^{-25} \, \mathrm{kg}$) at $B=B_{\mathrm{res}}=543.25\,\mathrm{G}$~\cite{grimm,lithium}. Since it is an exceptionally narrow (closed-channel dominated) resonance for a fermionic species, it enables one to create, through an adiabatic sweep of the magnetic field, a system of almost bare molecules, very weakly hybridized with the atomic sectors. The resonance appears in the scattering of atoms in the internal states labelled as the $a$ and $b$ channels, which, in the high-field regime~\cite{grimm}, safely applicable at $B_{\mathrm{res}}$, coincide with
\begin{align}
\ket{\uparrow} \equiv \ket{ i_z=1, s_z=-1/2 }, \label{statea} \\
\ket{\downarrow} \equiv \ket{ i_z=0, s_z=-1/2 }, \label{stateb}
\end{align}
where $i_z$ and $s_z$ are the components along the magnetic field of the atomic nuclear and electronic spin, respectively, in units of $\hbar$. Henceforth, we will adopt the notation
\begin{equation}
\bm{S}=\bm{s}_1+\bm{s}_2, \qquad \bm{I}=\bm{i}_1+\bm{i}_2
\end{equation}
for the total spins.

The atom pair state~$\ket{ \bm{p}  + \bm{K}/2 \uparrow, - \bm{p}  + \bm{K}/2 \downarrow }$ can be decomposed into the product
of the center-of-mass state~$\ket{ \bm{K}_{\mathrm{cm}} }$ and an antisymmetric relative-motion state~$\ket{ \bm{p}\uparrow,-\bm{p}\downarrow}$, defined in~\eqref{eq:relmotdef}, which reads, in terms of the orbital and internal states, 
\begin{equation}
\ket{ \bm{p}\uparrow,-\bm{p}\downarrow  } =
\frac{1}{\sqrt{2}} \big( \ket{ \bm{p},-\bm{p}  }_{-} \otimes
\ket{\psi_{ab}^+}+ \ket{ \bm{p},-\bm{p}
}_{+}  \otimes \ket{\psi_{ab}^-} \big),
\end{equation}
where
\begin{eqnarray}
\ket{\psi_{ab}^{\pm}} &=& \frac{1}{\sqrt{2}} \big( \ket{ \uparrow} \otimes
\ket{\downarrow } \pm \ket{ \downarrow} \otimes\ket{ \uparrow } \big), 
\nonumber\\
\ket{ \bm{p},-\bm{p}  }_{\pm} &=& \frac{1}{\sqrt{2}} \big( \ket{  \bm{p}} \otimes \ket{-\bm{p} } \pm \ket{- \bm{p}} \otimes \ket{\bm{p} } \big).
\end{eqnarray}

Notice that, while the pseudospin state can be either symmetric or antisymmetric, the atom pair is always an {\it electronic spin} triplet state, corresponding to $S_z=-1$. The resonant bound state is an $s$-wave, electronic spin singlet state, whose nuclear spin is a quasi-degenerate mixture of $I=0$ and $I=2$~\cite{grimm}. The state $\ket{\psi_{\mathrm{M},\bm{K}}}$ of the molecule can thus be  factorized into
\begin{equation}\label{eq:bound}
\ket{\psi_{\mathrm{M},\bm{K}}} = \ket{ \bm{K}_{\mathrm{cm}} }
\otimes \ket{ \phi_{\mathrm{M}} } \otimes \ket{\Phi_I} \otimes
\ket{ S=0, S_z=0 },
\end{equation}
where $\ket{ \phi_{\mathrm{M}} }$ is  the spherically symmetric orbital wavefunction and $\ket{\Phi_I}$ is the superposition of the $I=0,2$ states coupled to the atom pair state by $H_{\mathrm{int}}$. The open and closed channels are connected by the hyperfine coupling~\cite{AMcoherence1} between the nuclear and electronic spins of each atom
\begin{equation}
H_{\mathrm{int}} \equiv H_{\mathrm{hf}} = A_{\mathrm{hf}} \left(
\bm{i}_1 \cdot \bm{s}_1 + \bm{i}_2 \cdot \bm{s}_2 \right),
\end{equation}
where the constant $A_{\mathrm{hf}}$ can be deduced from the
hyperfine splitting between the states with total spin $f=3/2$ and
$f=1/2$ at zero magnetic field. In the case of $^6$Li, the
hyperfine splitting reads $\delta E_{\mathrm{hf}} = 228 \,
\mathrm{MHz} \,\hbar$, and its relation with the coupling constant
is $A_{\mathrm{hf}}=(2/3) \delta E_{\mathrm{hf}}$.

Since the hyperfine Hamiltonian does not act on the the relative motion degrees of freedom, only the term involving $\ket{ \bm{p},-\bm{p}
}_{+}$ in the pair wavefunction is coupled to the $s$-wave bound state and contributes to $G(\bm{p})$ in~\eqref{eq:amcoupling}. The action of the hyperfine Hamiltonian on the related pseudospin state $\ket{\psi_{ab}^-}$ reads
\begin{eqnarray}\label{eq:hyperfine}
& & \frac{1}{A_{\mathrm{hf}}} H_{\mathrm{hf}} \ket{\psi_{ab}^-}= 
-\frac{1}{2} \ket{\psi_{ab}^-} + \frac{1}{2} \ket{ I=1, I_z=0 }
\otimes \ket{ \tau } \nonumber \\ &  & \quad + \frac{\sqrt{3}}{6} \Bigl(
\ket{ I=2, I_z=0 }  
- 2
\sqrt{2} \ket{ I=0, I_z=0 } \Bigr) \otimes \ket{ \sigma },
\nonumber\\
\end{eqnarray}
with $\ket{\tau}\equiv \ket{S=1,S_z=0}$ and $\ket{\sigma}\equiv\ket{S=0,S_z=0}$. Assuming that the initial molecule condensate is obtained from free atoms by adiabatically sweeping the magnetic field across the resonance, the bound nuclear state in~\eqref{eq:bound} can be identified with the normalized state
\begin{equation}
\ket{ \Phi_I } = \frac{1}{3} \ket{ I=2, I_z=0 } - \frac{2
\sqrt{2}}{3} \ket{ I=0, I_z=0 }
\end{equation}
associated with the electronic spin singlet~$\ket{ \sigma }$ in~\eqref{eq:hyperfine}. Based on these considerations, the form factor of the atom-molecule
interaction in~\eqref{eq:HAMdef} reads
\begin{eqnarray}
G(\bm{p}) & = & \frac{1}{\sqrt{2}} \langle \phi_{\mathrm{M}} \ket{
\bm{p},-\bm{p} }_+ \bigl( \bra{\sigma}\otimes\bra{\Phi_I}
\bigr) H_{\mathrm{hf}} \ket{\psi_{ab}^-} \nonumber \\ & = &
\frac{\sqrt{3}}{2} A_{\mathrm{hf}}
\frac{\tilde{\phi}_{\mathrm{M}}(\bm{p})}{\sqrt{V}},
\end{eqnarray}
with $V$ the quantization volume, $\phi_{\mathrm{M}}(\bm{r})=\langle \bm{r}
\ket{\phi_{\mathrm{M}}}$ and
\begin{equation}\label{eq:phitilde}
\tilde{\phi}_{\mathrm{M}}(\bm{p}) = \int \dd\bm{r} \,
\phi_{\mathrm{M}}(\bm{r}) \ee^{\ii \bm{p} \cdot \bm{r} / \hbar}.
\end{equation}
To determine an explicit expression for the Fourier-transformed molecular orbital wavefunction~$\tilde{\phi}_{\mathrm{M}}$, we shall consider in the following section an approximated interatomic potential.

\section{Morse potential approximation}\label{sec:morse}

The resonant molecular state is the last excited vibrational
state, with $v=38$, of the interatomic potential in the electronic
spin singlet state. Since its wave function is spherically
symmetric, a radial wave function $\chi(r)$, with $r=|\bm{r}|$, can be used,
\begin{equation}
\phi_{\mathrm{M}}(\bm{r})  =
\frac{\chi(r)}{r}.
\end{equation}
The radial equation can be solved if the interatomic potential,
characterized by a repulsive core and an attractive tail, is
approximated by a Morse potential
\begin{equation}\label{eq:morse}
V(r)= D \left( \ee^{-2 \alpha (r-r_0)} - 2 \ee^{- \alpha (r-r_0)}
\right),
\end{equation}
which depends on three parameters: the absolute value of the minimum $D$, the radial distance $r_0$ of the minimum and the inverse length constant $\alpha$. The discrete negative energy levels are characterized by an integer vibrational quantum number $v$, ranging from zero to $\lfloor\lambda\rfloor$, 
with
\begin{equation}
\lambda = \frac{\sqrt{2 \mu D}}{\hbar \alpha},
\end{equation}
where $\mu$ is the reduced mass, and read
\begin{equation}
\epsilon(v) = - \frac{(\hbar \alpha)^2}{2 \mu} \left( \lambda- v -
\frac{1}{2} \right)^2.
\end{equation}
The independent parameters $\lambda$, $\alpha$ and $r_0$ can be
fixed by fitting the relevant features of the physical state, namely the
potential depth $V_{\min}/\hbar= 250\,$THz, the dissociation energy $E_{38}/\hbar\simeq 1.6 \,$GHz, the potential minimum position
$r_0 \simeq 5a_0$, with $a_0=5.29 \, 10^{-11} \, \mathrm{m}$ the Bohr
radius, and the classical turning points $a_{\min}\simeq 3a_0$ and $a_{\max}\simeq 43 a_0$. We choose the three
parameters of the Morse potential by preserving the following
features of the measured potential: the ratio of
the binding energy to the potential depth, the
extension of the classical motion for $v=38$ and the position of the inner turning
point. This ensures an accurate
reproduction of the spatial distribution of the state. The ratio $E_{38}/V_{\min}$
fixes the parameter $\lambda$ by imposing
\begin{equation}
\frac{|\epsilon(38)|}{D} = \frac{1}{\lambda^2} \left( \lambda- 38
- \frac{1}{2} \right)^2 = \frac{E_{38}}{V_{\min}} \quad
\Rightarrow \quad \lambda \simeq 38.6,
\end{equation}
while $\alpha$ is determined $E_{38}$ and $\delta r=a_{\max}-a_{\min}$ through
\begin{equation}
\alpha = \frac{1}{\delta r} \log \left(
\frac{1+\sqrt{1-E_{38}/V_{\min}}}{1-\sqrt{1-E_{38}/V_{\min}}}
\right) \simeq \frac{1}{3 a_0}.
\end{equation}
The minimum of the approximate potential is placed at
\begin{equation}
r_0 = r_{\min} + \frac{1}{\alpha} \log \left( 1+\sqrt{1-
\frac{E_{38}}{V_{\min}}} \right) \simeq 5.08 \, a_0,
\end{equation}
remarkably close to the measured value. Since $\lfloor\lambda\rfloor=38$, the considered resonant state is the most excited discrete level in the Morse potential, as in the real case.

\begin{figure}
\centering
\includegraphics[width=0.49\textwidth]{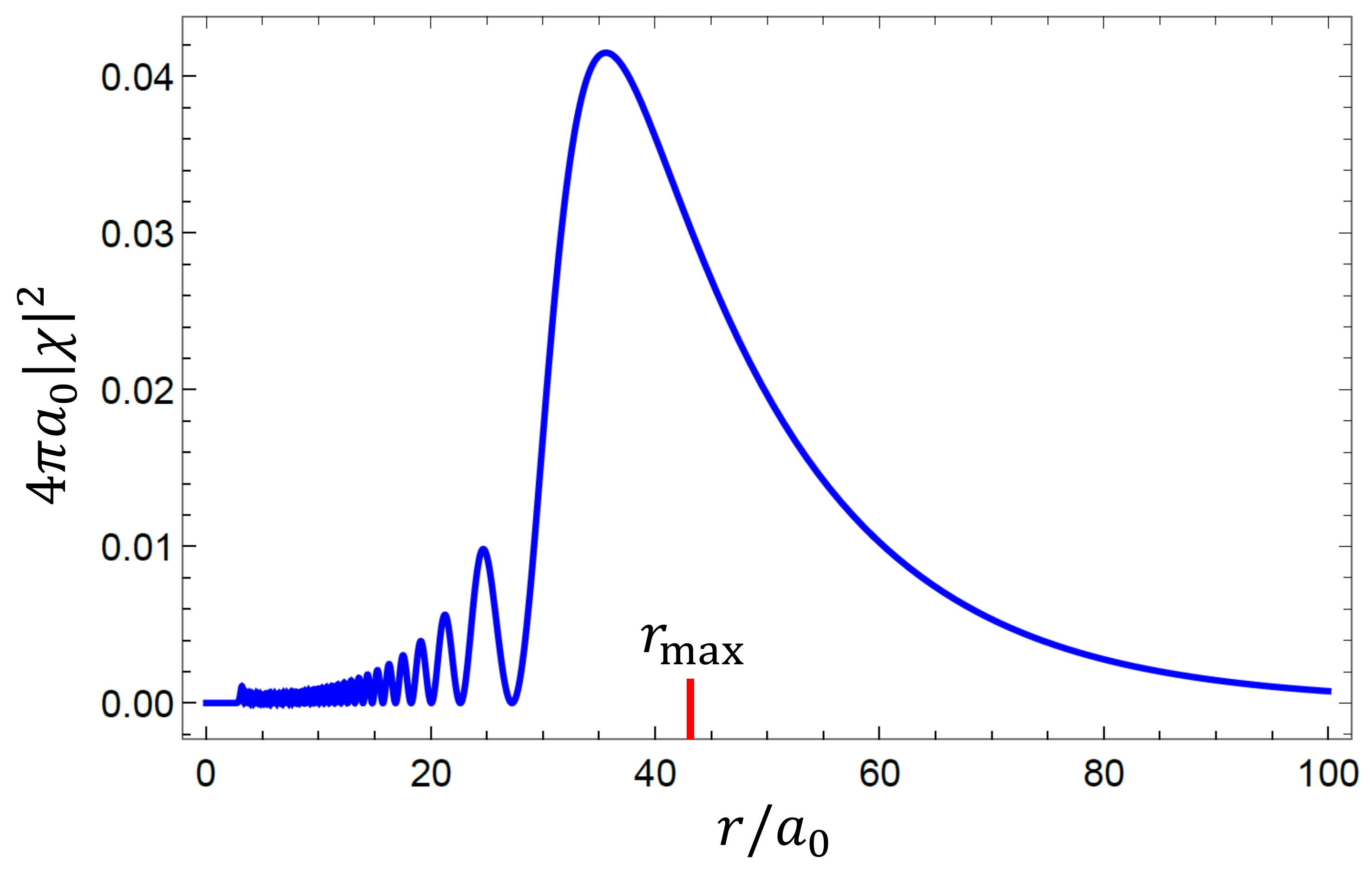}
\caption{Morse approximation of the radial probability density $4\pi|\chi(r)|^2$. 
The (red) tick represents the outer turning point of the classical motion at energy $E_{38}$.}\label{fig:wf}
\end{figure}

We can determine the wavefunction $\psi_{38}$, depending on the
adimensional variable $x=\alpha r$, as
\begin{eqnarray}\label{eq:wavef}
\psi_{38}(x) & = & N_{\lambda,38} (y(x))^{\lambda-77/2}
\ee^{-y(x)/2} L_{38}^{2\lambda -77} (y(x)), \nonumber \\
N_{\lambda,38} & = & \sqrt{ \frac{38! (2\lambda - 77) }{ \Gamma(2
\lambda- 38 ) } }, \nonumber
\\ y(x) & = & 2 \lambda \ee^{-(x- \alpha r_0)},
\end{eqnarray}
where $\Gamma(\beta)$ is the Euler gamma function and
$L_{\gamma}^{\delta}(z)$ is a generalized Laguerre polynomial, satisfying
\begin{equation}
\int_{0}^{\infty} \dd x\, |\psi_{38}(x)|^2 = 1.
\end{equation}
To obtain the correct normalization of the molecular wave
function $\phi_{\mathrm{M}}(\bm{r})$, the radial function
$\chi(r)$ must be related to~\eqref{eq:wavef} by
\begin{equation}
\chi (r) = \sqrt{ \frac{\alpha}{4 \pi} } \psi_{38}(y(\alpha r)).
\end{equation}
The radial wave function, plotted in Fig.~\ref{fig:wf}, is
evidently concentrated around the outer classical turning point,
and is significantly extended beyond it. The Fourier transform~\eqref{eq:phitilde} of the molecular wave
function, once the spherical symmetry of
$\phi_{\mathrm{M}}(\bm{r})$ is exploited and the $V\to\infty$
limit is taken, can be conveniently expressed in terms of an
integral involving the adimensional wave function~\eqref{eq:wavef},
\begin{equation}
\tilde{\phi}_{\mathrm{M}} (\bm{p}) = \frac{4 \pi \hbar}{p}
\int_0^{\infty} \dd r \, \chi(r) \sin\left( \frac{p r}{\hbar} \right)
= \sqrt{ \frac{4\pi}{\alpha} } \frac{\hbar}{p} \, F \left(
\frac{p}{\hbar\alpha} \right),
\end{equation}
with
\begin{equation}\label{eq:fp}
F(P) \equiv \int_0^{\infty} \dd x \, \psi_{38}(z(x)) \sin (Px).
\end{equation}
The behavior of the form factor $G(\bm{p})$, which is proportional
to $\tilde{\phi}_{\mathrm{M}} (\bm{p})$, is determined by the
adimensional function $F(P)/P$. At low momenta, the form factor
approaches a constant, since
\begin{equation}
\frac{F(P)}{P} \to  I_1 \equiv \int_0^{\infty} \dd x \, \psi_{38}(z(x)) x
\simeq 94.
\end{equation}
The square of $F(P)/P$ falls off rapidly with $P$, and is fitted with very good approximation by a Gaussian function $I_1 \exp (-(\alpha b P)^2)$, where $b\simeq 46.7 \, a_0$ can be
interpreted as a cutoff length.

\section{Self energy and  dynamics}\label{sec:self}

The survival probability amplitude of the initial one-molecule state (\ref{eq:instate}),
$\ket{\psi_{\mathrm{M},0}} = \b_{\bm 0}^{\dagger} \ket{0}$, reads for $t>0$
\begin{equation}
\label{eq:evolution}
\mathcal{A}(t) = \bra{\psi_{\mathrm{M},0}} \ee^{-\ii H t/\hbar}
\ket{\psi_{\mathrm{M},0}} = \frac{\ii}{2 \pi} \int_B \dd E \, \mathcal{G}(E)
\ee^{-\ii E t / \hbar}
\end{equation}
where the propagator
\begin{equation}\label{eq:prop}
\mathcal{G}(E)= \bra{\psi_{\mathrm{M},0}} \frac{1}{E-H}
\ket{\psi_{\mathrm{M},0}}
\end{equation}
is the expectation value of the resolvent $(E-H)^{-1}$ in the state $\ket{\psi_{\mathrm{M},0}}$, and the integration (Bromwich) path $B$ is a horizontal line in the complex energy upper half-plane, $\operatorname{Im} E >0$, with constant imaginary part. The propagator can
be expressed as
\begin{equation}
\mathcal{G}(E) = \frac{1}{E-E_{\mathrm{B}} - \Sigma(E)},
\end{equation}
with $E_{\mathrm{B}}$ the binding energy of the bare molecule, and $\Sigma(E)$ the self-energy,
representing all possible transitions generated by the interaction Hamiltonian $H_{\mathrm{hf}}$, and connecting $\ket{\psi_{\mathrm{M},0}}$ with itself, without involving $\ket{\psi_{\mathrm{M},0}}$ as an intermediate state.

The self-energy can always be expressed as an integral involving a
spectral function $\kappa(E)$, as 
\begin{equation}\label{eq:selfint}
\Sigma(E)= \int_{E_0}^{\infty} \dd E' \frac{ \kappa(E') }{ E-E' },
\end{equation}
for $E\in \mathbb{C}\setminus [E_0, +\infty)$,  with $E_0$ the ground energy of $H$. If one neglects, as a first approximation, scattering between free atoms
(namely, $U(\bm{p}) = 0$ in Eq.\ (\ref{eq:quartic})), the exact expression of the spectral function~\cite{cohentannoudji} reads
\begin{equation}
\kappa^0 (E) = \sum_{\bm{p}} |\bra{  \bm{p}\uparrow,
-\bm{p}\downarrow  } \hat{H}_{\mathrm{AM}}
\ket{\psi_{\mathrm{M},0}} |^2 \delta (E - E_{\bm{p}}),
\end{equation}
with $E_{\bm{p}}=p^2/m$ the energy of a pair of atoms with
mass $m$ and opposite momenta $\bm{p}$ and $-\bm{p}$. After a straightforward
manipulation, one obtains
\begin{eqnarray}
\kappa^0 (E) & = & \sum_{\bm{p}} | G(\bm{p}) |^2 \delta (E -
E_{\bm{p}}) \nonumber \\ & = & \frac{3
A_{\mathrm{hf}}^2}{4\pi\hbar\alpha} \sqrt{\frac{m}{E}} F^2\,\left(
\frac{\sqrt{m E}}{\hbar\alpha} \right) \theta(E).
\end{eqnarray}
It is evident that $\kappa^0(E) = 0$ if $E$ does not belong to the continuous spectrum of the free-fermion Hamiltonian.
Based on the expression~\eqref{eq:fp}, the spectral function also reads
\begin{equation}
\kappa^0(E) = \sqrt{E} f(E) \theta(E),
\end{equation}
where $f(E)$ is a single-valued function, analytic on the whole complex plane. Thus, at low energy, the
functional form of the spectral function is determined only by the
density of states. Let us recall that the spectral function is
related by the Fermi golden rule to the approximated inverse
lifetime $\gamma_{\mathrm{GR}}(E_{\mathrm{B}}) = 2\pi
\kappa^0(E_{\mathrm{B}})/\hbar$ of the molecular state with binding
energy $E_{\mathrm{B}}$, which is nonzero  only when
$E_{\mathrm{B}}$ is positive. The integral~\eqref{eq:selfint},
providing the self-energy $\Sigma^0$, is well defined on the whole
complex plane except for the positive real axis, on which a branch cut is present~\cite{cohentannoudji}.

\begin{figure*}
\centering
\includegraphics[width=0.95\textwidth]{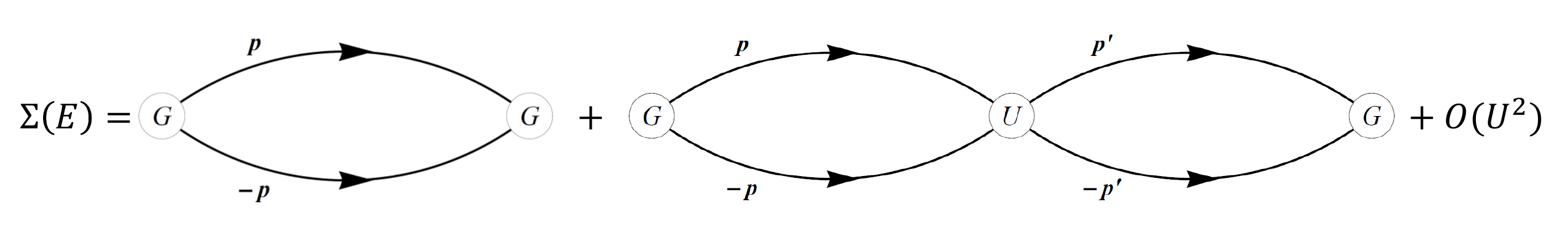}
\caption{Feynman diagrams of the contributions to the self-energy $\Sigma(E)$ of the molecular state up to the second order in $U$. Each pair of lines represents a free atom-pair propagator $E-p^2/m$, while the functions $G$ and $U$ must be computed taking into account the momenta of the propagators connected to the specific vertex. Momenta with different names represent different integration variables.}\label{fig:selfenergy}
\end{figure*}

Interatomic scattering renormalizes the two-atom propagator
appearing in the self-energy of the molecule~\cite{bosefermi1}, introducing new processes that change the momenta of the emitted atom pair before recombination. Let us analyze how these processes, diagrammatically represented in Fig.~\ref{fig:selfenergy}, affect the self-energy. Consider for
simplicity the case in which the coupling between the atom pair and the molecule is constant
\begin{equation}
G \equiv G(0) = \frac{ \sqrt{3}}{2 \sqrt{V}} A_{\mathrm{hf}}
\tilde{\phi}_{\mathrm{M}} (0) = \sqrt{\frac{3\pi}{V}} I_1
\frac{A_{\mathrm{hf}}}{\alpha^{3/2}},
\end{equation}
as well as the interatomic coupling $U\equiv U(0)$, physically related to the background scattering length
$a_{\mathrm{bg}}\simeq 60 a_0$ (namely, the scattering length far from the Feshbach resonance) by~\cite{pricoupenko}
\begin{equation}
U = \frac{U_c}{V} \left( 1- \frac{\sqrt{\pi} b}{a_{\mathrm{bg}}}
\right)^{-1}, \quad \text{with } U_c \equiv - \frac{ 4 \pi^{3/2}
\hbar^2 b}{m}.
\end{equation}
In the last equality, $V$ is the normalization volume and $b$ is the characteristic length of a
Gaussian cutoff function, which has been used to regularize
$\mathrm{O}(a_{\mathrm{bg}}^2)$ terms, the first-order term being
the usual result $4\pi\hbar^2 a_{\mathrm{bg}}/m$. We will effectively assume that the form factor of interatomic scattering has the same cutoff length as the atom-molecule transition
(see e.g.\ Ref.~\cite{pricoupenko}). Incidentally, let us remark that, since $U>0$, no BCS transition
is expected at low temperature.

The self-energy in absence of scattering reads
\begin{equation}
\Sigma^{(0)} (E) = \Sigma^0(E) = \sum_{\bm{p}}
\frac{G^2}{E-E_{\bm{p}}},
\end{equation}
while the first-order term in $U$ is
\begin{equation}
\Sigma^{(1)} (E) = \sum_{\bm{p}} \frac{G}{E-E_{\bm{p}}} U
\sum_{\bm{p}'} \frac{G}{E-E_{\bm{p}'}} = \frac{U}{G^2} (\Sigma^0
(E))^2.
\end{equation}
Notice that, since $G\propto V^{-1/2}$ and $U\propto V^{-1}$, the ratio $U/G^2$ is independent of the normalization volume $V$. Since the $n$-th order term in $U$ reads
\begin{equation}
\Sigma^{(n)} (E) = \Sigma^0(E) \left( \frac{U}{G^2} \Sigma^0 (E)
\right)^n ,
\end{equation}
it is possible to sum all the contributions to the self-energy and get
\begin{equation}\label{eq:sigmau}
\Sigma (E) = \sum_{n=0}^{\infty} \Sigma^{(n)} (E) =
\frac{\Sigma^0(E)}{1-\frac{U}{G^2}\Sigma^0(E)}.
\end{equation}
The divergent $\Sigma^0(E)$ is regularized by the replacement $G
\to G(\bm{p})$. The complete spectral function can be obtained
using the identity $\operatorname{Im} \Sigma(E+\ii 0^+) = - \pi
\kappa(E)$, which yields
\begin{equation}
\kappa(E) = \frac{ \kappa^0(E) }{ \left[ 1 - \frac{U}{G^2}
\fint_{0}^{\infty} dE' \frac{\kappa^0(E')}{E-E'}
\right]^2 + \left[ \frac{U}{G^2} \pi \kappa^0(E) \right]^2 },
\end{equation}
with $\fint$ denoting principal value integration. The new function $\kappa(E)$ inherits the square-root singularity of
$\kappa^0(E)$.

To determine the decay rate, one performs the analytic continuation of the self
energy from the upper to the lower complex half-plane, through the branch
cut. The continuation results in the following expression for the
self-energy on the second Riemann sheet:
\begin{equation}
\Sigma_{\mathrm{II}}(E) = \Sigma(E) - 2\pi \ii \kappa(E), \qquad \operatorname{Im}E<0,
\label{eq:SigamII}
\end{equation}
with $\kappa(E)$ the analytic continuation of the spectral function from the positive
real axis. The decay rate and wavefunction renormalization are determined by the pole of the
propagator~\eqref{eq:prop} in the lower half-plane of the second Riemann sheet, which satisfies
\begin{equation}\label{eq:pole}
E_{\mathrm{pole}} = E_{\mathrm{B}} +
\Sigma_{\mathrm{II}}(E_{\mathrm{pole}})
\end{equation}
and can be considered as a function of the binding energy, that, close to the resonance, is approximately linear in the magnetic field,
\begin{equation}
E_{\mathrm{B}}(B) = 2 \mu_{\mathrm{Bohr}} (B - B_0),
\end{equation}
with $\mu_{\mathrm{Bohr}}=9.27 \, 10^{-24} \, \mathrm{J/T}$ the
Bohr magneton, and $B_0$ the value at which the bare binding energy is equal to the continuum threshold of the atom pair. The linear dependence is due to the Zeeman splitting
between the bound state, which is an electronic singlet, and the
free state, characterized by $S_z=-1$ [see Eqs.~\eqref{statea}-\eqref{stateb}]. Once the pole has been
determined, the survival amplitude in the initial state can be split as
\begin{equation}\label{eq:evopole}
\mathcal{A}(t) = Z \ee^{- \ii (E_{\mathrm{B}} + \Delta
E) t/\hbar} \ee^{- \gamma  t/2} +
\tilde{\mathcal{A}}(t),
\end{equation}
where the  first term, giving an exponential decay, is related to the integration around the pole, while the second term $\tilde{\mathcal{A}}$ contains residual contributions from branch cut integrations, and is responsible for all deviations from the exponential law~\cite{strev,cohentannoudji}. The energy shift $\Delta E$ and the  decay rate $\gamma$  of the bound state are related to the real and imaginary parts of the pole by
\begin{equation}
\Delta E 
= \operatorname{Re} E_{\mathrm{pole}}, 
\qquad \gamma 
= - \frac{2}{\hbar}  \operatorname{Im} E_{\mathrm{pole}}. 
\label{eq:decayrate}
\end{equation}
The amplitude factor $Z$ is the residue at the pole, related to the first derivative of the self-energy by
\begin{equation}
\label{eq:zetaampl}
Z =  \left[  \frac{\dd G^{-1}}{\dd E}(E_{\mathrm{pole}} )\right]^{-1}
=
\frac{1}{1-\Sigma_{\mathrm{II}}'(E_{\mathrm{pole}}
)} = \frac{ \dd E_{\mathrm{pole}} (E_{\mathrm{B}}) }{\dd E_B},
\end{equation}
where the last equality follows from the pole definition~\eqref{eq:pole}.

\section{Poles of the propagator}\label{sec:poles}

The analytic properties of the propagator~\eqref{eq:prop}, discussed  in the previous section, provide the basis to characterize the stability properties of the molecular state with varying binding energy $E_B$. If the binding energy is significantly smaller
than zero, we expect the molecular state to be stable and very well approximated by $\ket{\psi_\mathrm{M,0}}$, while hybridization with the atomic sector would increase as the magnetic field approaches the resonance. When the molecule is stable, a real pole, representing the
energy of the dressed molecular state, can be found on the first
Riemann sheet, at an energy smaller than the branching point at $E=0$. Actually, real negative solutions of the equation
\begin{equation}
E_{\mathrm{pole}} = E_{\mathrm{B}}(B) + \Sigma(E_{\mathrm{pole}})
\label{eq:negativepole}
\end{equation}
are found up to the value $B_{\mathrm{res}}$ of magnetic field such that
\begin{equation}
E_{\mathrm{B}}(B_{\mathrm{res}}) + \Sigma(0) = 0 \quad \Rightarrow
\quad B_{\mathrm{res}}-B_0 = 0.208 \, \mathrm{G}.
\end{equation}
As $B$ approaches $B_{\mathrm{res}}$ from below,
$\Sigma'(E_{\mathrm{pole}})$ diverges due to the singularity of
the spectral function in $E_0$. Thus the derivative $Z(E_B)=\dd E_{\mathrm{pole}}(E_{\mathrm{B}})/ \dd E$, defined as in \eqref{eq:zetaampl} on the second Riemann sheet, vanishes. Therefore, approaching the resonance from below, $E_{\mathrm{pole}}$ admits the quadratic approximation
\begin{equation}
E_{\mathrm{pole}}(E_{\mathrm{B}}) \sim 2 \mu_{\mathrm{B}}^2 \eta
(B-B_{\mathrm{res}})^2 ,
\end{equation}
as $B\uparrow B_{\mathrm{res}}$,
with
\begin{equation}
\eta \equiv \left. \frac{\dd Z(E_B) }{\dd E_{\mathrm{B}}} \right|_{B\uparrow B_{\mathrm{res}}} =
\lim_{\epsilon\uparrow 0}
\frac{\Sigma''(\epsilon)}{(1-\Sigma'(\epsilon))^3} ,
\end{equation}
while it is linear far from the resonance. The very small width found for the pure
quadratic region ($\lesssim 10^{-6}\,\mathrm{G}$) is expected from
the extreme narrowness of the resonance. The value
$B_{\mathrm{res}}$ represents the actual position of the
resonance, which is an experimental parameter and not a prediction of the model.
Close to the resonance, the energy of the bound state is related
to the positively diverging scattering length $a(B)$ by~\cite{grimm}
\begin{equation}
|E_{\mathrm{pole}}| \simeq \frac{\hbar^2}{m a^2(B)} \quad
\Rightarrow \quad a^2(B) \simeq \frac{ \hbar^2}{ 2
\mu_{\mathrm{B}}^2 |\eta| m (B-B_{\mathrm{res}})^2}.
\end{equation}
Comparing this relation with the general expression~\cite{lithium1995,grimm}
\begin{equation}
a(B)= a_{\mathrm{bg}} \left( 1- \frac{\Delta}{B-B_{\mathrm{res}}}
\right),
\end{equation}
of the resonant scattering length, one obtains the result $\Delta \simeq 0.08\,\mathrm{G}$, which is
very close to the measured value~\cite{lithium}.

\begin{figure}
\centering
\includegraphics[width=0.48\textwidth]{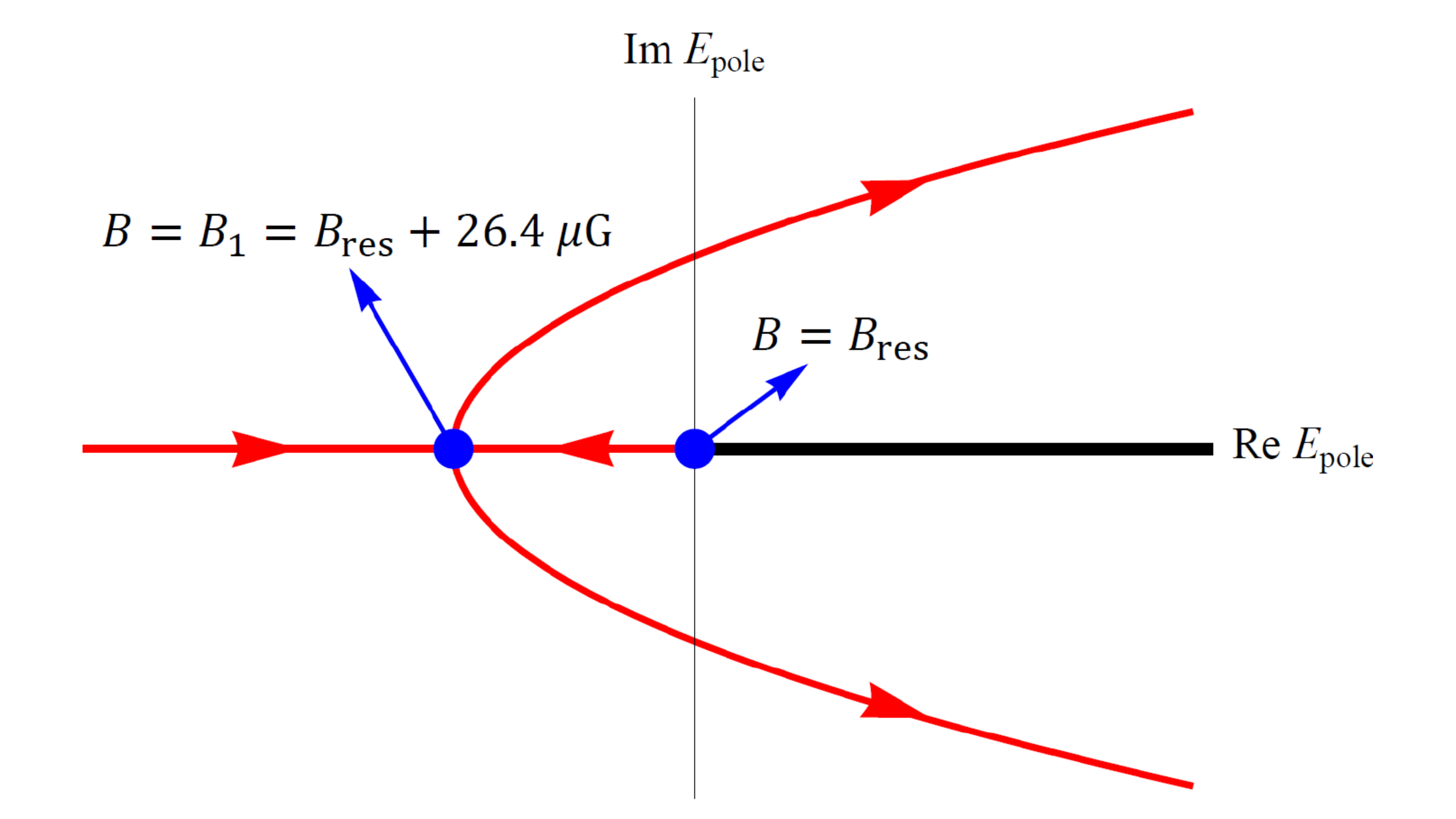}
\caption{Locus of the poles of the propagator of the initial state $\ket{\psi_{\mathrm{M},0}}$ in the second Riemann sheet. The poles in the lower half-plane ($\operatorname{Im}E<0$) are solution of the equation $E_{\mathrm{pole}}=E_{B}+\Sigma (E_{\mathrm{pole}}) + 2\pi \ii \kappa(E_{\mathrm{pole}})$, while the poles in the upper half-plane ($\operatorname{Im}E>0$) satisfy $E_{\mathrm{pole}}=E_{B}+\Sigma (E_{\mathrm{pole}}) - 2\pi \ii \kappa(E_{\mathrm{pole}})$. The two equations coincide for real negative poles. The second pole on the negative real axis originates from a negative pole on the first Riemann sheet which passes through the origin for $B=B_{\mathrm{res}}$. The two real poles collide and bounce off into two complex conjugate poles for $B=B_1 = B_{\mathrm{res}} +
26.4 \,\mu\mathrm{G}$. Arrows on the red lines indicate the motion of poles as the magnetic field increases.}\label{fig:poles}
\end{figure}

The zeros of
\begin{equation}
\mathcal{G}_{\mathrm{II}}(E)^{-1}=E-E_{\mathrm{B}}-\Sigma_{\mathrm{II}}(E),
\end{equation} 
with $\operatorname{Im}E<0$, are the poles of the propagator on the second Riemann
sheet, which are relevant for the decay dynamics. The time-reversed
process is instead related to the zeros of
$\mathcal{G}_{\mathrm{II}}(E)^{-1}$ in the upper half-plane, $\operatorname{Im}E>0$,
where one gets 
\begin{equation}
\Sigma_{\mathrm{II}} (E) = \Sigma(E) + 2\pi \ii \kappa(E),
\label{eq:SigamII2}
\end{equation}
which coincides with the analytical continuation of $\Sigma(E)$ from below the branch
cut on the first Riemann sheet. Indeed, due to the square-root singularity
of the spectral function, the two expressions~\eqref{eq:SigamII} and~\eqref{eq:SigamII2} of $\Sigma_{\mathrm{II}}$ coincide on the negative real axis, since
$ \sqrt{x+\ii 0^+} = - \sqrt{x-\ii 0^+}$ for $x<0$. 

On the negative axis of the second Riemann sheet a real pole with $Z>0$,
solution of 
\begin{equation}
E_{\mathrm{pole}} = E_{\mathrm{B}}(B) + \Sigma(E_{\mathrm{pole}}) + 2\pi \ii \kappa(E_{\mathrm{pole}}+\ii 0^+),
\end{equation}
exists for all $B < B_{\mathrm{res}}$, and moves towards the origin as $B$ increases. 
As we have seen above, also  the pole on the negative real axis of the
first Riemann sheet~\eqref{eq:negativepole} moves towards the origin as $B$ increases.
At $B = B_{\mathrm{res}}$ it hits the branch point at the origin and bounces back on the
second Riemann sheet moving backwards on the negative real axis as $B$ increases. The two real poles on the second Riemann sheet collide at $B=B_1$ with
\begin{equation}
B_1 \simeq B_{\mathrm{res}} + 2.64\times 10^{-5} \, \mathrm{G}, 
\end{equation}
bouncing off each other and generating two complex conjugate poles, in the lower and upper half-planes. The locus of the poles
in the second Riemann sheet is shown in Fig.~\ref{fig:poles}, with the arrows pointing toward increasing magnetic field.
The real part of the complex poles increases linearly with $B-
B_1$ for all $B>B_1$, while their imaginary parts scale
approximately like $\sqrt{B- B_1}$.

\begin{figure}
\centering
\includegraphics[width=0.49\textwidth]{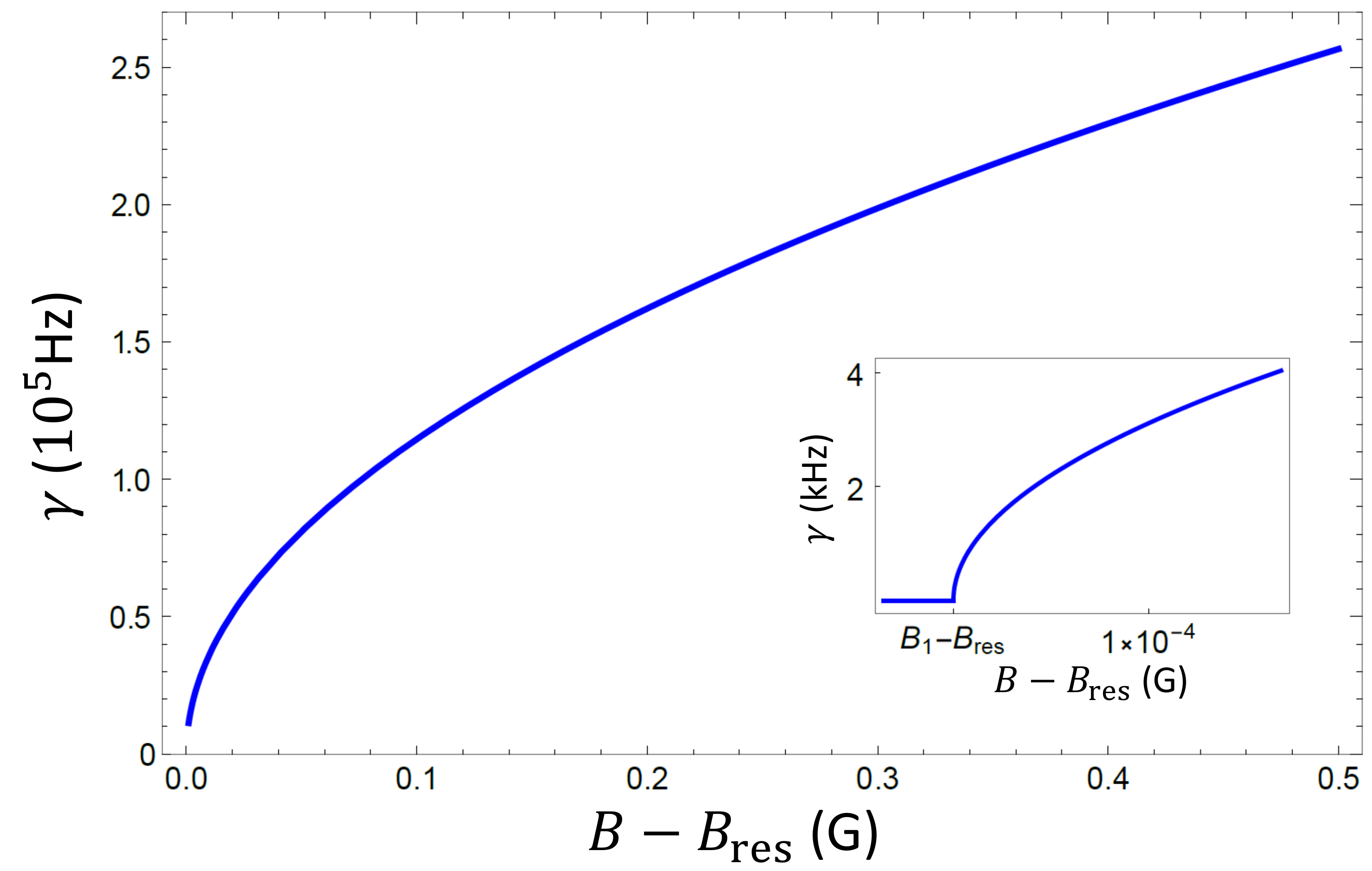}
\caption{Inverse lifetime $\gamma$ of the unstable molecule for $B>B_{\mathrm{res}}$. Inset: zoom on the region close to $B=B_{\mathrm{res}}$, where, even though the decay rate vanishes for $B_{\mathrm{res}}<B<B_1$, the molecule is still unstable.}\label{fig:gamma}
\end{figure}

The imaginary part of the complex pole in the lower half-plane yields the decay rate $\gamma$ of the unstable molecule through~\eqref{eq:decayrate},
which is plotted in Fig.~\ref{fig:gamma}. The lifetime is considerably shifted with respect to the golden rule result, obtained from the spectral density at $U=0$, since, close to the crossing ($B\sim B_0$), the atom-molecule coupling can no longer be treated as a perturbation.

\begin{figure}
\centering
\includegraphics[width=0.49\textwidth]{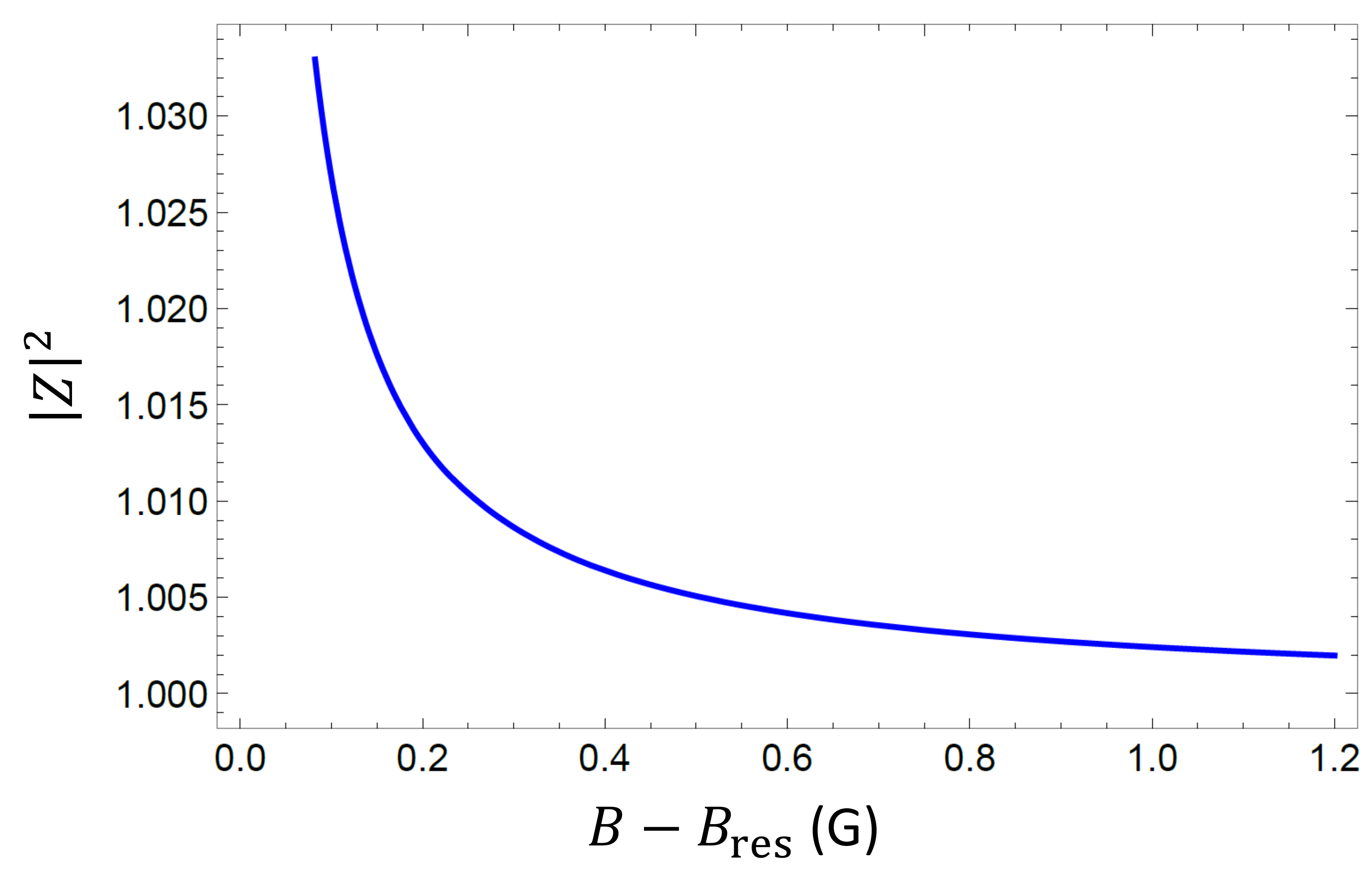}
\caption{Square modulus $|Z|^2$ of the wave function renormalization $Z$ in~\eqref{eq:evopole}. $|Z|^2$ coincides with
the extrapolated value at $t=0$ of the exponential part of the
molecule survival probability, and diverges
at $B=B_1>B_{\mathrm{res}}$ as $(B-B_1)^{-1/2}$.}\label{fig:zeta}
\end{figure}

The squared modulus of the wave function renormalization $Z$ in Eq.~\eqref{eq:evopole}
represents the extrapolation of the exponential part of the
survival probability to $t=0$,
\begin{equation}
P(t) = | \mathcal{A} (t) |^2 \simeq |Z|^2 \exp \left( - \gamma t
\right) .
\end{equation}
The plot in Fig.~\ref{fig:zeta} shows that for $B-B_{\mathrm{res}}
\lesssim 10^{-2} \, G$, the value of $|Z|^2$ is significantly
larger than one (which is only possible for unstable states \cite{heraclitus}), and
diverges as $B$ approaches $B_1$ from above like $(B-B_1)^{-1/2}$.

\section{Time evolution}\label{sec:time}

The increasing value of $|Z|^2$ as the system approaches the
resonance is a marker of the strong deviation of the 
molecular survival probability $P(t)$ from an exponential
law. Deviations are usually expected for very short times (Zeno
region), where the behavior of the survival probability for a finite-energy initial state
must be quadratic, and for very
long times, where for physical Hamiltonians a power-law tail supersedes  the vanishing
exponential part. In our system, a peculiar structure of the decay
laws emerges close to the resonance, which is mainly due to the
form of the spectral function, scaling like $\sqrt{E}$ for
practically the whole relevant energy range.

\begin{figure}
\centering \subfigure[]{\includegraphics[width=0.49\textwidth]{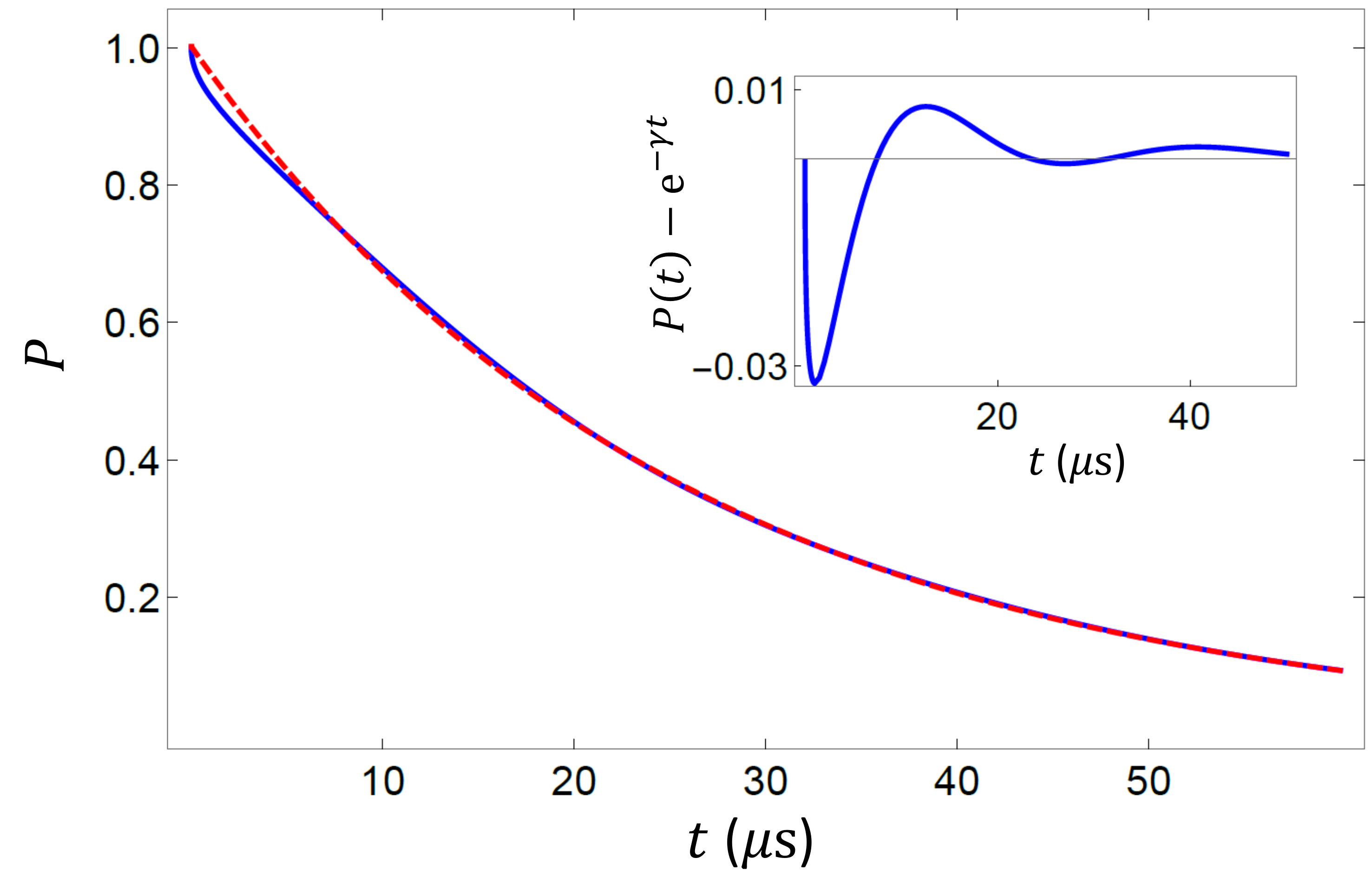}}
\subfigure[]{\includegraphics[width=0.49\textwidth]{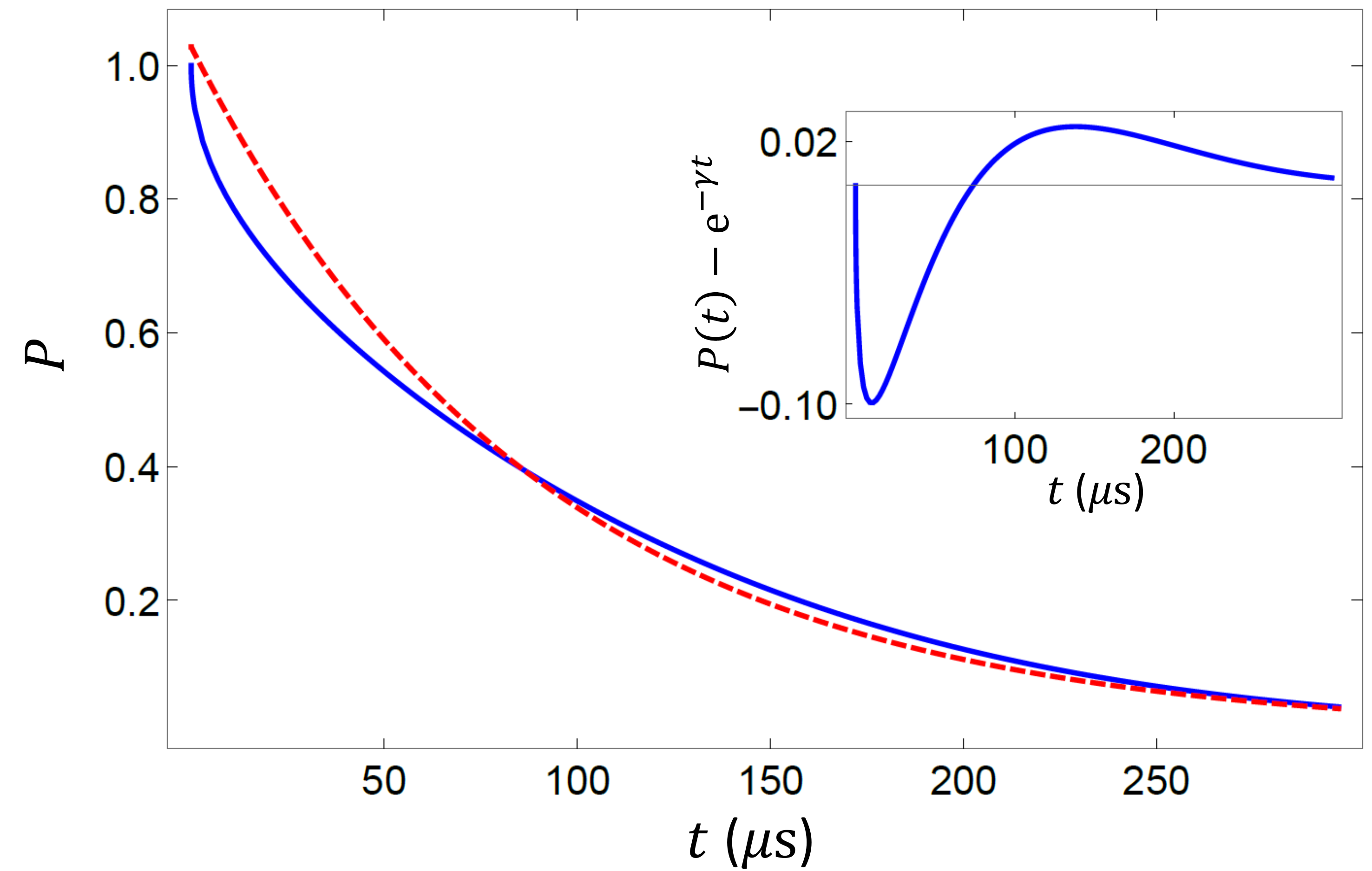}}
\caption{Evidence of nonexponential decay at short times for $B>B_1$ in the cases (a) $B-B_{\mathrm{res}}=12 \,\mathrm{mG}$ and (b) $B-B_{\mathrm{res}}=0.92 \,\mathrm{mG}$. In case (a), $\gamma=3.96\times 10^4\,\mathrm{s}^{-1}$ and $|Z|^2-1=2\times 10^{-3}$; in case (b), $\gamma=1.11\times 10^4\,\mathrm{s}^{-1}$ and $|Z|^2-1=2.8\times 10^{-2}$. The solid (blue) lines represent the exact survival probability $P(t)$ of the molecule, approaching the asymptotic dashed (red) curves $|Z|^2 \exp(-\gamma t)$ after a transient. If the magnetic field is very close to the resonance, nonexponentiality is enhanced, and the intersection point between the survival probability and the asymptotic exponential shifts towards $t\to\infty$. In the insets, the difference between the survival probability and a pure exponential curve with lifetime $1/\gamma$ is plotted. }
\label{fig:pt}
\end{figure}

\begin{figure}
\centering
\includegraphics[width=0.49\textwidth]{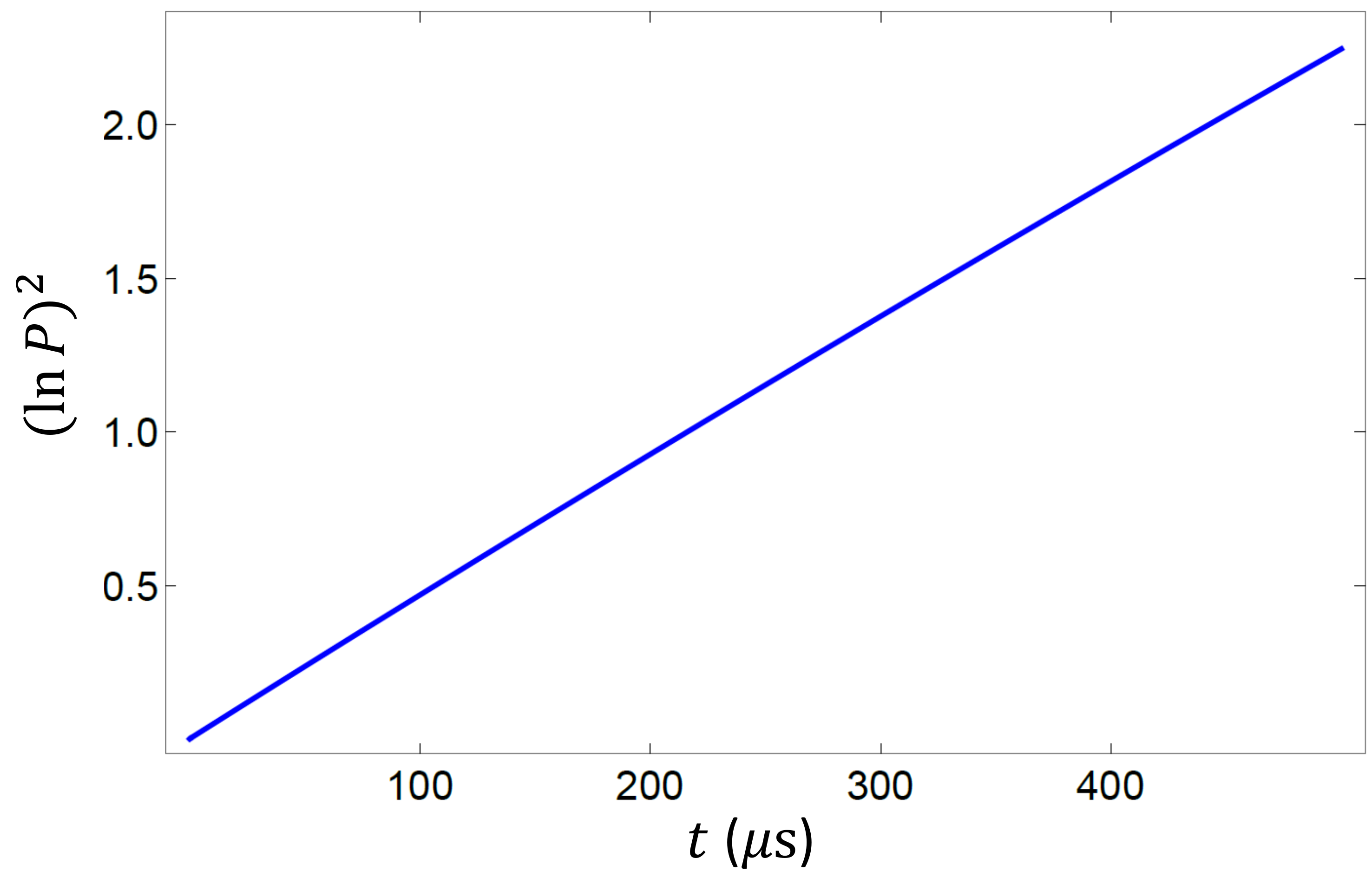}
\caption{Squared logarithm of the molecular survival probability for $B-B_{res} = 18 \,\mu\mathrm{G} $. In the intermediate region $B_{\mathrm{res}}<B<B_1$, in which the molecule is unstable but the decay rate is vanishing, the survival probability $P(t)$ follows with very good approximation a stretched exponential law $P(t)= \exp (- a \sqrt{t} )$. Forerunners of the stretched-exponential decay can be observed in the transient parts of decays with $\gamma \neq 0$, when $B$ approaches $B_1$.}\label{fig:pt02081}
\end{figure}

In order to analyze the effects we discussed close to resonance, and quantify the nonexponentiality of decay, let us compare the plots in Fig.~\ref{fig:pt}. In Fig.~\ref{fig:pt}(a), the computed survival probability approaches the
asymptotic curve $|Z|^2 \ee^{-\gamma t}$ after a time $\simeq 1/3\gamma$. Figure \ref{fig:pt}(b)
represents a situation in which the imaginary pole is very close to
the real axis, and $|Z|^2\simeq 1.03$. In this case, the
survival probability intersects the asymptotic exponential at a
time that is very close to $1/\gamma$, with the
survival probability already reduced by more than one half. As the magnetic field approaches $B_1$ from above, the first intersection between $P(t)$ and $|Z|^2 \ee^{-\gamma t}$ tends to infinity. Thus, in the limit $B\downarrow B_1$, the exponential regime is never reached and a new decay law emerges.

An analysis of time evolution of molecules in the intermediate
range $B_{\mathrm{res}}<B<B_1$ helps to characterize the emergent
decay law. Indeed, the square-logarithmic plot in Fig.~\ref{fig:pt02081} shows that, in this intermediate magnetic field range, decay is characterized by an approximate
stretched exponential law
\begin{equation}
P(t) \simeq \ee^{- a t^{\beta} },
\end{equation}
with $\beta$ very close to $1/2$, though slightly modulated in the considered time interval.

This law emerges as $B\downarrow B_1$ (see Fig.~\ref{fig:pt}) in the
first part of the decay, after the initial Zeno
region (unresolved in the plots, since the curvature $\tau_Z$ of the survival probability at the origin is of the order of $10\,\mathrm{ns}$), and gradually replaces the exponential law, which is
pushed towards infinity. Below $B_1$, the stretched exponential
law entirely dominates the relevant decay dynamics.

\section{Conclusions and outlook}\label{sec:conc}

We have analyzed the time evolution of an unstable Feshbach molecule decaying into a pair of fermionic atoms with opposite momenta. We have shown that, while the decay is exponential for magnetic fields far from the resonance, when the magnetic field approaches the resonance value, the decay is dominated by a stretched-exponential law $P(t)\simeq \exp(-a \sqrt{t})$. 

Stretched exponentials appear in the phenomenological description of a variety of physical phenomena, in classical statistical physics, glassy dynamics, and low-energy 1D Bose gases~\cite{stretched2,stretched1,Citro}. They are often used to describe relaxation in disordered or complex systems, when different local dynamics give rise to superpositions of simple exponential decays, whose average effect yields stretched exponential behaviour. 
The appearance of a stretched exponential relaxation in the context described in the present Article is therefore of interest, in that it bridges the gap between complexity and typical quantum relaxation in a cold gas.

We intend to devote future work to the study of the time evolution of cold atomic systems in presence of collective effects, such as macroscopic quantum tunneling in mixtures of Bose-Einstein condensates \cite{kyt}, confined in arbitrary potentials \cite{binary1,binary2}. It would be interesting to understand whether curious time evolutions like the one analyzed in this work are present in different quantum situations.

\section*{Acknowledgments}
PF and SP are partially supported by Istituto Nazionale di Fisica Nucleare (INFN) through the project ``QUANTUM". FVP is supported by INFN through the project ``PICS''.
PF is partially supported by the Italian National Group of Mathematical Physics (GNFM-INdAM).

\end{document}